\newcommand{\MyTitle}{The role of the probability current for time measurements}
\title{\MyTitle}
\newcommand{\MyAuthor}{Nicola Vona, Detlef Dürr}
\author{\MyAuthor}
\newcommand{\MySubject}{Time, Measurement, Quantum Mechanics, Bohmian Mechanics}
\newcommand{\MyPdfKeywords}{\MySubject}  

\documentclass[a4paper,11pt,final]{article}

\usepackage{microtype}   
\usepackage{natbib}

\usepackage[latin1]{inputenc} 
\usepackage{letltxmacro} 
\usepackage{ifdraft} 
\usepackage{color}    
\usepackage{graphicx}   
\usepackage{subfig}    
\usepackage{caption}    
\usepackage{ifthen}
\usepackage{versions} 

\newcommand{\question}[1]{\begin{center}\emph{#1}\end{center}}

\usepackage[ulem={normalem,normalbf}]{changes} 
\makeatletter
\@namedef{Changes@AuthorColor}{red}
\colorlet{Changes@Color}{red}
\makeatother

\usepackage[colorinlistoftodos,
			\ifoptionfinal{disable}{}
		]{todonotes} 

\LetLtxMacro\oldlistoftodos\listoftodos
\renewcommand{\listoftodos}{
\makeatletter
\let\oldchapter\chapter
\let\chapter\section
\protect\hypertarget{todolist}{} 
\oldlistoftodos
\let\chapter\oldchapter
\makeatother}

\LetLtxMacro\oldtodo\todo

\makeatletter
\newcommand{\smalllinkedtodo}[2][]{\oldtodo[caption={#2}, #1]{%
\if@todonotes@inlinenote
#2
\else
\renewcommand{\baselinestretch}{0.85}\selectfont\footnotesize#2\par
\fi
\hfill\hyperlink{todolist}{$\uparrow$}
}}
\makeatother

\providecommand{\comment}{}
\renewcommand{\comment}[2][]{\smalllinkedtodo[color= green,#1]{#2}}

\newcommand{\fodetleftitle}{\textbf{For Detlef: }}
\newcommand{\fordetlef}[2][]{\smalllinkedtodo[color= yellow,#1]{\fodetleftitle#2}}

\renewcommand{\todo}[2][]{\smalllinkedtodo[#1]{#2}} 

\definecolor{defaultmarkcolor}{rgb}{0,0.5,0}
\renewcommand{\markversion}[2][defaultmarkcolor]
	{
	\ifoptionfinal {\includeversion{#2}} {	
	 \newenvironment{#2}
			{\bigskip\bigskip\bigskip
				\todo[inline,color=#1,caption={Begin #2}]{\color{white}Begin #2}
					\color{#1!50!black} 
				 }
			{
				\todo[inline,color=#1,nolist
						]{\color{white}End #2}
				\bigskip\bigskip}
	}}

\usepackage{hyperref}   
\providecommand{\MyTitle}{}
\providecommand{\MyAuthor}{}
\providecommand{\MySubject}{}
\providecommand{\MyPdfKeywords}{} 
\hypersetup{colorlinks, citecolor=black, filecolor=black, linkcolor=black, urlcolor=black, 
bookmarksopen=true, 
bookmarksnumbered=true, 
pdftitle={\MyTitle}, pdfauthor={\MyAuthor}, pdfsubject={\MySubject}, pdfkeywords={\MyPdfKeywords} 
}
\usepackage{amsmath}
\usepackage{amssymb}
\usepackage{mathrsfs}   
\usepackage{braket}    
\usepackage{empheq}  
\let\tempepsilon\varepsilon
\let\varepsilon\epsilon
\let\epsilon\tempepsilon
\newcommand{\R}{\mathbb{R}}    
\newcommand{\PP}{\mathbb{P}}   
\newcommand{\abs}[1]{\lvert#1\rvert}  
\newcommand{\Abs}[1]{\left \lvert#1\right \rvert}  
\newcommand{\of}[1]{\mathchoice{{\scriptstyle(#1)}}{{\scriptstyle(#1)}}{{\scriptscriptstyle(#1)}}{{\scriptscriptstyle(#1)}}}   
\newcommand{\norm} [1]{\| #1 \|}

\newcommand{\1}{\mathrm{1}}  
\newcommand{\povm}{\textsc{povm}}

\newcommand{\de}{\mathrm{d}}    
\newcommand{\Der}[2]{\frac{\de^{#1}\phantom{#2}}{\de {#2}^{#1}}}    
\newcommand{\der}[3]{\frac{\de^{#1}#2}{\de {#3}^{#1}}}    
\newcommand{\intinde}[1]{\int \! \de #1}  
\newcommand{\intdef}[2]{\int_{\scriptscriptstyle#1}^{\scriptscriptstyle#2}\!\! }   
\newcommand{\intdefde}[3]{\intdef{#1}{#2} \de #3 \  }   

\usepackage[T1]{fontenc}
\usepackage{fourier}

\ifoptionfinal	{\excludeversion{VersionPERSONALNOTES}}  
				{\markversion{VersionPERSONALNOTES}}

\begin{document}
\maketitle

\listoftodos
\tableofcontents


\newpage

\section{Introduction}
Think of a very simple experiment, in which a particle is sent towards a detector.
\question{When will the detector click?}
Imagine to repeat the experiment many times, starting a stopwatch at every run.
The instant at which the particle hits the detector will be different each time, forming a statistics of \emph{arrival times}.
Experiments of this kind are routinely performed in almost any laboratory, and are the basis of many common techniques, collectively known as time-of-flight methods (\textsc{tof}).
In spite of that, how to theoretically  describe an arrival time measurement is a very debated topic since the early days of quantum mechanics \citep{Pauli1958}.
It is legitimate to wonder why it is so easy to speak about a position measurement at a fixed time, and so hard to speak about a time measurement at a fixed position.
An overview of the main attempts and a discussion of the several difficulties they involve can be found in \citep{MugaLeavens2000,MugaMayatoEgusquiza2008,MugaMayatoEgusquiza2009}.

In the following, we will discuss the theoretical description of time measurements with particular emphasis on the role of the probability current.

\section{What is a measurement?}
We will start recalling the general description of a measurement in quantum mechanics in terms of \emph{positive operator valued measures} (\povm s).
This framework is less common than the one based on self-adjoint operators, but is more general and more explicit than the latter.

\subsection{Linear measurements -- \povm s}\label{sec:Povms}
When we speak about a \emph{measurement}, what are we speaking about?\\
A measurement is a situation in which a physical system of interest interacts with a second physical system, the apparatus, that is used to inquire into the former.
In general, we are  interested in those cases in which the experimental procedure is fixed and independent of the state of the system to be measured given as input; these cases are called \emph{linear measurements}.
The meaning of this name will be clarified in the following.
The analysis of the general properties of a linear measurement, and of the general mathematical description of such a process, has been carried out mostly by \citet{Ludwig1983a}, and finds a natural completion within Bohmian mechanics \citep{DuerrGoldsteinZanghi2004}.
In the following, we will present a simplified form of this analysis \citep{DurrTeufel2009,NielsenChuang2000}.

We will denote by $x$ the configuration of the system and by $\psi_0$ its initial state, element of the Hilbert space $L^2(\R^{3n})$, while we will use  $y$ for the configuration of the apparatus and $\Phi_0\in L^2(\R^{3N})$ for  its ready state; moreover, we will denote by $(0,T)$ the interval during which the interaction constituting the measurement takes place.
The evolution of the composite system is a usual quantum process, so the state at time $T$ is 
\begin{equation}
\Psi_T \of{x,y}  =  (U_T\,  \Psi_0)\of{x,y}  =  U_T\, (\psi_0\,\Phi_0) \of{x,y}   ,
\end{equation}
where $U_T$ is a  unitary operator on $L^2(\R^{3(N+n)})$.
We call such an interaction a \emph{measurement} if for every initial state $\psi_0$ it is possible to write the final state $\Psi_T$ as
\begin{equation}
\Psi_T \of{x,y}  =  \sum_\alpha  \psi_\alpha\of x \, \Phi_\alpha\of y  ,
\end{equation}
with the states $\Phi_\alpha$ normalized and clearly distinguishable, i.e.\ with supports $G_\alpha = \{ y \,|\, \Phi_\alpha\of y \neq 0 \}$ macroscopically separated.
This means that after the interaction it is enough to ``look'' at the position of the apparatus pointer to know the state of the apparatus.
Each support $G_\alpha$ corresponds to a different result of the experiment, that we will denote by  $\lambda_\alpha$.
One can imagine each support to have a label with the value $\lambda_\alpha$ written on it: if the position of the pointer at the end of the measurement is inside the region $G_\alpha$, then the result of the experiment is $\lambda_\alpha$.
The probability of getting the outcome $\lambda_\alpha$ is
\begin{equation}\label{eq:ProbAlpha}
\mathbb{P}_\alpha 
	= \intinde{x} \intdefde{G_\alpha}{}{y}  | \Psi_T\of{x,y} |^2
	= \intinde{x} \intdefde{G_\alpha}{}{y}  | \psi_\alpha\of x\,  \Phi_\alpha\of y |^2
	= \int   | \psi_\alpha\of{x} |^2\,  \de{x}  ,
\end{equation}
indeed $\Phi_{\alpha'}\of{y} =0$ $\forall y\in G_\alpha$, $\alpha'\neq \alpha$, and  the $\Phi_\alpha$ are normalized.
Consider now the projectors $P_\alpha$ that act on the Hilbert space  $L^2(\R^{3(N+n)})$
 of the composite system  and project to the subspace {$L^2(\R^{3n}\times G_\alpha)$}  corresponding to the pointer in the position $\alpha$, i.e.\ in particular
\begin{equation}
P_\alpha \Psi_T =  \psi_\alpha\,  \Phi_\alpha  .
\end{equation}
Through the projectors $P_\alpha$ we can  define the operators $R_\alpha$ such that
\begin{equation}
P_\alpha \Psi_T =  \psi_\alpha\,  \Phi_\alpha  =  (R_\alpha \psi_0) \; \Phi_\alpha  ,
\end{equation}
that means $R_\alpha \psi_0 = \psi_\alpha$.
Finally, we can also define the operators $O_\alpha =  R_\alpha^\dag R_\alpha$.
These operators are directly connected to the probability \eqref{eq:ProbAlpha} of getting the outcome $\alpha$
\begin{equation}\label{eq:ProbO}
\PP_\alpha = \norm{\psi_\alpha}^2  = \braket{\psi_0 | O_\alpha \psi_0} .
\end{equation}
Therefore, \emph{the operators $O_\alpha$ together with the set of  values $\lambda_\alpha$ are sufficient to determine any statistical quantity related to the experiment}.
The fact that any experiment of the kind we have considered can be completely described by a set of linear operators, explains the origin of the name \emph{linear measurement}.
Equation\eqref{eq:ProbO} implies also that the operators $O_\alpha$ are \emph{positive}, i.e.\
\begin{equation}
\braket{\psi_0 | O_\alpha \psi_0}  \geq 0
	\qquad \forall \psi_0 \in L^2(\R^{3n})  .
\end{equation}
In addition, they constitute a \emph{decomposition of the unity}, i.e.\ 
\begin{equation}
\sum_\alpha O_\alpha = \1  ,
\end{equation}
as a consequence of the unitarity of $U_T$ and of the orthonormality of the states $\Phi_\alpha$, that imply
\begin{equation}
1 = \norm{\psi_0\Phi_0}^2  
	= \norm{\Psi_T}^2  
	= \sum_\alpha \norm{\psi_\alpha}^2
	= \sum_\alpha \braket{\psi_0 | O_\alpha \psi_0}
	\qquad \forall \psi_0 \in L^2(\R^{3n})  .
\end{equation}
A set of operators with these features is called discrete \emph{positive operator valued measure}, or simply \povm.  It is a measure on the discrete set of values $\lambda_\alpha$. In case the value set is a continuum, the \povm\  is a Borel-measure on that continuum, taking values in the set of positive linear operators.

It is important to note that in the derivation of the \povm{} structure the orthonormality of the states $\Phi_\alpha$ and the unitarity of the overall evolution play a crucial role, while in general, the states $\psi_\alpha$ do not need to be neither orthogonal nor distinct.

In case the operators $O_\alpha$ happen to be orthogonal projectors, then the usual measurement formalism of standard quantum mechanics is recovered by defining  the selfadjoint operator $\hat A = \sum_\alpha \lambda_\alpha \, O_\alpha$.
Physically, this condition is achieved \comment{It is not also necessary as well as sufficient because it is true also for predictable measurements (introduced by Landau and Peierls).} for example in a reproducible measurement, i.e.\ one in which the repetition of the measurement using the final state $\psi_\alpha$ as input, gives the result $\alpha$ with certainty.

We remark that calculating the action of a \povm{} on a given initial state requires that that state is evolved for the duration of the measurement together with an apparatus, and therefore its evolution in general differs from the evolution of the system alone.
This circumstance is evident if one thinks that the state of the system after the measurement will depend on the measurement outcome.%
\footnote{It will be an eigenstate of the selfadjoint operator corresponding to the measurement, in case it exists.}
Usually, if the measurement is not explicitly modeled, this evolution is considered as a black box that takes a state as input and gives an outcome and another state as output.
It is important to keep in mind that the measurement formalism always entails such a departure from the autonomous evolution of the system, even if not explicitly described.

\subsection{Not only \povm s\label{sec:NotPOVMs}}
Although a linear measurement is a very general process, there are many quantities that are not measurable in this sense.
An easy example is the probability distribution of the position $|\psi|^2$.
Indeed, suppose to have a device that shows the result $\lambda_1$ if the input is a particle in a state for which the position is distributed according to $|\psi_1|^2$, and $\lambda_2$ if it is in a state with distribution $|\psi_2|^2$.
If the process is described by a \povm, the linearity of the latter requires that when the state $\psi_1+\psi_2$ is given as input, the result is \emph{either $\lambda_1$ or $\lambda_2$}, as for example the result of a measurement of spin on the state $\ket{\mathrm{up}}+\ket{\mathrm{down}}$ is either ``up'' or ``down''.
On the contrary, if the device was supposed to measure the probability distribution of the position, the result had to be $\lambda_+$, corresponding to $|\psi_1 + \psi_2|^2$,  possibly distinct both from $\lambda_1$ and from $\lambda_2$.

To overcome a limitation of this kind, the only possibility is to give up on linearity, accepting as measurement also processes different than the one devised in the previous section.
These processes use additional information about the $x$-system, for example giving a result dependent on previous runs, or adjusting the interaction according to the state of the $x$-system.
In particular, to measure the probability distribution of the position one exploits the fact that $|\psi\of{x}|^2 = \braket{\psi | O_x | \psi}$, where $O_x = \ket x \bra x$ is the density of the \povm{} corresponding to a position measurement.
Instead of measuring directly $|\psi|^2$, one measures $x$, and repeats the measurement on many systems prepared in the same state $\psi$.
The distribution $|\psi|^2$ is then recovered from the statistics of the results of the position measurements.
The additional information needed in this case is that all the $x$-systems used as input were prepared in the same state.
The outcome shown by the apparatus depends then on the preparation procedure of the input state: if we change it, we have to notify the change to the apparatus, that needs to know how to collect together the single results to build the right statistics.

For other physical quantities not linearly measurable, like for example the wave function, a similar, but more refined strategy is required.
This strategy is known as \emph{weak measurement} \citep{AharonovAlbertVaidman1988}.
An apparatus to perform a weak measurement is characterized first of all by having a very weak interaction with the $x$-system; loosely speaking, we can say that the states $\psi_\alpha$ are very close to the initial state $\psi_0$.
As a consequence of such a small disturbance, the information conveyed to the $y$-system by the interaction is very little.
The departure from linearity is realized in a way similar to that of the measurement of $|\psi|^2$: the single run does not produce any useful information because of the weak coupling, therefore the experiment is repeated many times on many $x$-systems prepared in the same initial state $\psi_0$; the result of the experiment is recovered from a statistical analysis of the collected data.

The advantage of this arrangement is that the output state $\psi_\alpha$ can be used as input for a following  linear measurement of usual kind (\emph{strong}), whose reaction is almost as if its input state was directly $\psi_0$.
In this case the experiment yields a joint statistics for the two measurements, and it is especially interesting to \emph{postselect} on the value of the strong measurement, i.e.\ to arrange the data in sets depending on the result of the strong measurement and to look at the statistics of the outcomes of the weak measurement inside each class.
For example, a weak measurement of position followed by a strong measurement of momentum, postselected on the value zero for the momentum, allows to measure the wave function \citep{LundeenSutherlandPatel2011}.

The nonlinear character of weak measurements becomes apparent if one understands the many repetitions they involve in terms of a calibration.
Indeed, one can think of the last run as the actual measurement, and of all the previous runs as a way for the apparatus to collect information about the $x$-system used in the last run, profiting from the knowledge that it was prepared exactly as the $x$-systems of the previous runs. 
The $x$-systems used in the preliminary phase can be then considered part of the apparatus, used to build the joint statistics needed to decide which outcome to attribute to the last strong measurement.
For example, the result of the experiment could be the average of the previous weak measurements postselected  on the strong value obtained in the last run.
If we then change the initial wave function $\psi_0$ to some $\psi\mathrlap{\smash{'}}_0$, the calibration procedure has to be repeated.
In this case, the apparatus itself depends on the state of the $x$-system to be measured, breaking linearity.

\section{Time statistics}
Now we finally come to our topic: time measurements.
At first, we have to note that there are several different experiments that can be called time measurements: measurements of dwell times, sojourn times, and so on.
We will refer in the present discussion exclusively to \emph{arrival times}, although it is possible to recast everything to fit any other kind of time measurement.
More precisely, we will consider the situation described at the beginning: a particle is prepared in a certain initial state and a stopwatch is set to zero; the particle is left evolving in presence of a detector at a fixed position; the stopwatch is read when the detector clicks.
The time read on the stopwatch is what we call \emph{arrival time}.

A measurement of this kind is necessarily linear, and we can ask for the statistics of its outcomes given the initial state of the particle.
If, for example, we measure the position at the fixed time $t$, then we can predict the statistics of the results by calculating the quantity
\begin{equation}
\braket{ \psi_t | x }  \braket{ x | \psi_t }  .
\end{equation}
Which calculation do we have to perform to predict the statistics of the stopwatch readings with the detector at a fixed position?

\subsection{The semiclassical approach}
\fordetlef{New section; no changes made elsewhere.}
Arrival time measurements are routinely performed in actual experiments, and they are normally treated semiclassically: essentially, they are interpreted as momentum measurements.
The identification with momentum measurements is motivated by the fact that the detector is at a distance $L$ from the source usually much bigger than the uncertainty on the initial position of the particles, so one can assume that each particle covers the same length $L$.
Hence, the randomness of the arrival time must be a consequence of the uncertainty on the momentum, and the time statistics must be given by the momentum statistics.
For a free particle in one dimension, the connection between  time and momentum is provided by the classical relation $p\of{t}  = m \,L/t$.
By a change of variable, this relation implies that the probability density of an arrival at time $t$ is
\begin{equation}\label{eq:TraditionalDensity}
|\tilde\psi\of{p\of{t}}|^2\, \Abs{\der{}{p\of{t}}{t}}  
=  |\tilde\psi\of{p\of{t}}|^2\, \frac{m L}{t^2}   ,
\end{equation}
where $\tilde\psi$ is the Fourier transform of the wave function $\psi$.

This semiclassical approach is justified by the distance $L$ being very big, that is true for most experiments so far performed.
On the other hand, we tacitly assumed that the particle moves on a straight line with constant velocity $v$, whose ignorance is the source of the arrival time randomness: such a classical picture is inadequate to describe the behavior of a quantum particle in general conditions, and is expected to fail in future, near-field experiments.
A deeper analysis is needed.

\subsection{An easy but false derivation}\label{sec:EasyDerivation}

Consider that the particle crosses the detector at time $t$ with certainty.
This implies that the particle is on one side of the detector before $t$, and on the other side after $t$. 
One can therefore think that it is possible to connect the statistics of arrival time to the probability that the particle is on one side of the detector at different times. Because the latter is known, this seems like a good strategy.

For simplicity we will consider only the one dimensional case, that already entails all the relevant features that we want to discuss.%
\footnote{The same treatment is possible in three dimensions, provided that the detector is sensitive only to the arrival time and not to the arrival position, and that the detecting surface divides the whole space in two separate regions (i.e.\ it is a closed surface or it is unbounded).}
The detector is located at the origin; we will assume the evolution of the particle in presence of the detector to be very close to that of the particle alone.
We consider the easiest possible case: a free particle, initially placed on the negative half-line and moving towards the origin, i.e.\ prepared in a state $\psi_0$ such that
\begin{align}
\psi_0\of x &\approx 0  \qquad  \forall x\geq0; \label{eq:NegativeX}
\\
\tilde\psi\of p &=0  \qquad  \forall p\leq0  ,
\end{align}
where $\tilde\psi$ denotes the Fourier transform of $\psi_0$, and eq.\ \eqref{eq:NegativeX} is a shorthand for $\intdef{0}{\infty}\psi_0\of x \de x \ll 1$.
The particle can only have positive momentum, therefore it will get at some time to the right of the origin and thus it has to cross the detector from the left to the right.

One might think that the probability to have a crossing at a time $\tau$ later than $t$ is equal to the probability that the particle at $t$ is still in the left region,
\begin{equation}\label{eq:LeftProb}
\PP (\tau\geq t)
= \PP (x\leq 0;t)
= \intdefde{-\infty}{0}{x}  \abs{\psi_t\of{x}}^2  .
\end{equation}
Conversely, the probability that the particle arrived at the detector position before $t$ is
\begin{equation}
\PP (\tau<t)
= 1 - \PP (\tau\geq t)
= \intdefde{0}{\infty}{x}  \abs{\psi_t\of{x}}^2  .
\end{equation}
Therefore, the probability density $\Pi\of{t}$ of a crossing at $t$ is
\begin{equation}
\Pi\of t = \Der{}{t}  \PP (\tau<t)
= \intdefde{0}{\infty}{x} \partial_t \abs{\psi_t\of{x}}^2  .
\end{equation}
We can now make use of the continuity equation for the probability
\begin{equation}
\partial_t\, \bigl(|\psi_t\of x |^2\bigr) +  \partial_x  j\of{ x, t}  = 0  ,
\end{equation}
that is a consequence of the Schrödinger equation, with the probability current
\begin{equation}\label{eq:J}
j\of{ x, t} \coloneqq  \tfrac{\hbar}{m}\,  \Im \psi_t^*\of x\ \partial_x\psi_t\of x  .
\end{equation}
Substituting,
\begin{equation}\label{eq:QFlux1D}
\Pi\of{t}
=
-\intdefde{0}{\infty}{x} \partial_x  j\of{ x, t} 
=
 j\of{ x=0, t}   .
\end{equation}
Thus,  the probability density $\Pi\of{t}$ of an arrival at the detector at time $t$ is equal to the probability current $j\of{ x=0, t}$, \emph{provided everything so far has been correct}. Well, it hasn't. Eq.\ \eqref{eq:LeftProb} is problematical. It is only correct if the right hand side is a monotonously decreasing function of time, or, equivalently, if the current in \eqref{eq:QFlux1D} is always positive. But that is in general not the case and it is most certainly not guaranteed by asking that the momentum be positive. Indeed, even considering only free motion and positive momentum, there are states for which the current is not always positive, a circumstance  known as \emph{backflow} (for an example, see the appendix).
But a probability distribution must necessarily be positive, hence, \emph{the current can not be equal to the statistics of the results of any linear measurement}, i.e.\ there is no \povm{} with density $O_t$ such that
\begin{equation}
\braket{ \psi_0 | O_t | \psi_0 } = j\of{ x=0, t}  .
\end{equation}

This problem is well known \citep{Allcock1969b} and has given rise to a long debate, aiming at finding a quantum prediction for the arrival time distribution with the needed \povm{} structure \citep{MugaLeavens2000}.
%

One might wonder: How can it be that the momentum is only positive, and yet the probability that the particle is in the left region is not necessarily decreasing?
A state with only positive momentum is such that, if we \emph{measure} the momentum, then we find a positive value with certainty.
This is not the same as saying that the particle \emph{moves} only from the left to the right when we do not measure it.
Actually, in strict quantum-mechanical terms, it does not even make sense to speak about the momentum of the particle when it is not measured, as it does not make sense to speak about its position if we do not measure it, and therefore there is no way of conceiving how the particle moves in this framework.
Think for example of a double slit setting: we can speak about the position of the particle at the screen, but we can not say through which slit the particle went.

Although the quantum-mechanical momentum is only positive, the conclusion that the particle  moves only once from the left to the right is unwarranted.
Even more: it simply does not mean anything.

\subsection{The moral}
The problem with the simple derivation of the arrival time statistics is quite instructive, indeed it forces us to face the fact that quantum mechanics is really about measurement outcomes, and therefore it is a mistake to think of quantum-mechanical quantities as of quantities intrinsic to the system under study and independent of the measurement apparatus.

\section{The Bohmian view}
Bohmian mechanics is a theory of the quantum phenomena alternative to quantum mechanics, but giving the same empirical predictions \citep[see][]{DurrTeufel2009,DurrGoldsteinZanghi2013}.
The two theories share at their foundation the Schrödinger equation.
Quantum mechanics complements it by some further axioms like the collapse postulate, and describes all the objects around us only in terms of wave functions.
On the contrary, according to Bohmian mechanics the world around us is composed by actual point particles moving on continuous paths, that are determined by  the  wave function.
The Schrödinger equation is in this case supplemented by a guiding equation that specifies the relation between the wave function and the motion of the particles.
The usual quantum mechanical formalism is recovered in Bohmian mechanics as an effective description of measurement situations \citep[see][]{DurrTeufel2009}.

The main difference between quantum and Bohmian mechanics is that the first one is concerned only with measurement outcomes, while the second one gives account of the \emph{physical reality} in any situation.
Although every linear experiment corresponds to a \povm{} according to quantum mechanics as well as to Bohmian mechanics \citep{DuerrGoldsteinZanghi2004}, for the former \povm s are the fundamental objects the theory is all about, while for the latter they are only very convenient tools that occur when the theory is used to make predictions.

We saw already how interpreting quantum-mechanical quantities as intrinsic properties of a system is mistaken, and how the framework of quantum mechanics is limited to measurement outcomes.
In Bohmian mechanics the particle has a definite trajectory, so it makes perfectly sense to speak about its position or velocity also when they are not measured, and it is perfectly meaningful to argue about the way the particle \emph{moves}.
In doing so, one has just to mind the difference between the outcomes of hypothetical  (quantum) measurements, and actual (Bohmian) quantities.

\subsection{The easy derivation again\dots}
Let's review the derivation of section \ref{sec:EasyDerivation} from the point of view of Bohmian mechanics.

To find out the arrival time of a Bohmian particle it is sufficient to literally follow its motion and to register the instant when it actually arrives at the detector position.
A Bohmian trajectory $Q\of t$ is determined by the wave function through the equation
\begin{equation}\label{eq:Bohm}
\dot Q \of t = \frac { j \of{Q\of t, t}  }    { | \psi\of{Q\of t, t}|^2 } ,
\end{equation}
with $j$  defined in eq.\ \eqref{eq:J}.
Hence, the Bohmian velocity, that is the actual velocity with which the Bohmian particle moves, is not directly related to the quantum-mechanical momentum, that rather encodes only information about the possible results of a hypothetical momentum measurement.
Even if the probability of finding a negative momentum in a measurement is  zero, the Bohmian particle can still  have negative velocity and arrive at the detector from behind, or even cross it more than once.%
\footnote{Note that the notion of \emph{multiple crossings} of the same trajectory is genuinely Bohmian, with no analog in quantum mechanics.}
It is in these cases that the current becomes negative.

\begin{figure}
\centering \includegraphics[width=.8\textwidth]{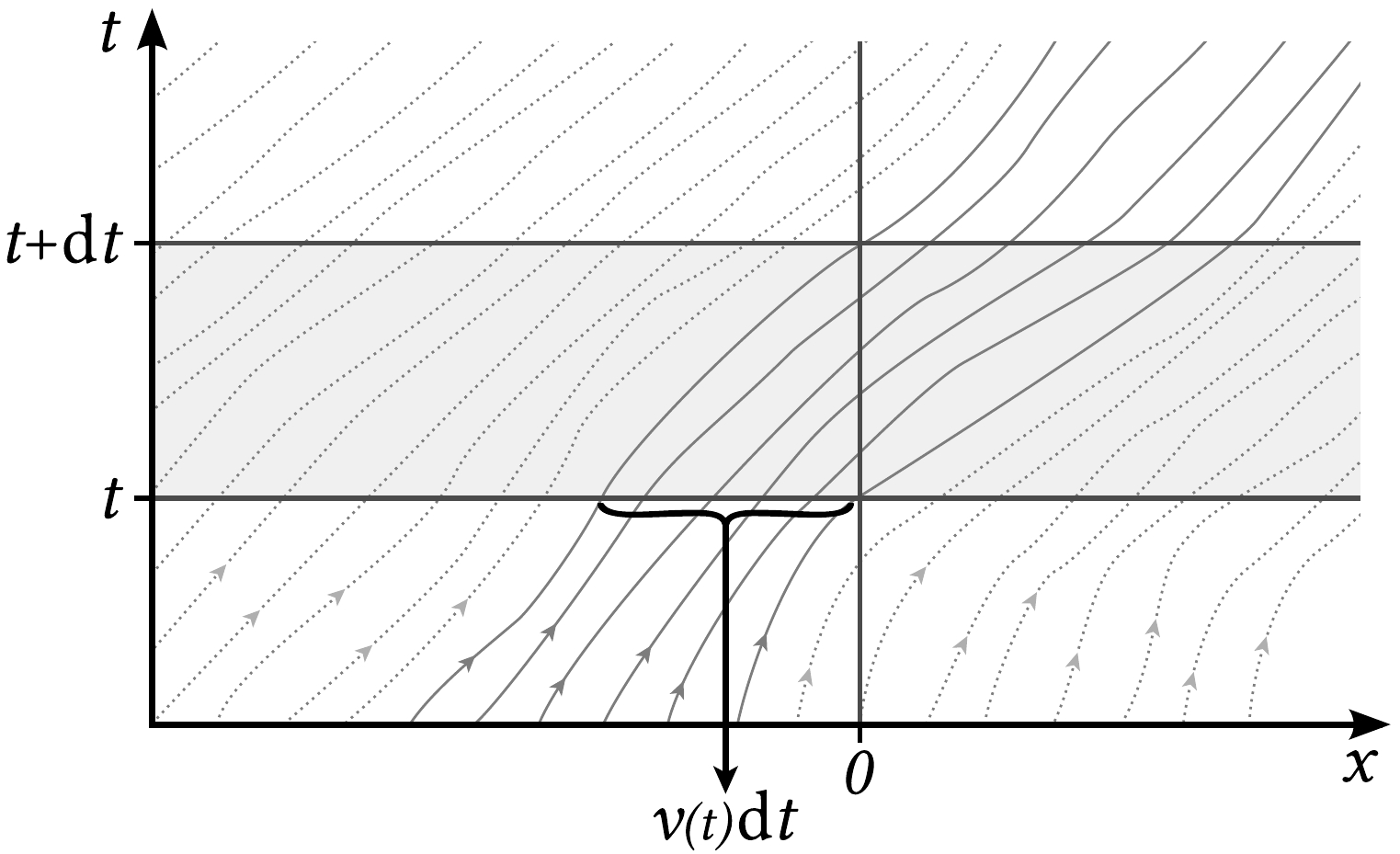}  
\caption{Bohmian trajectories in the vicinity of the detector, placed at $x=0$.
The trajectories, that cross the detector between the times $t$ and $t+\de t$, are those that at time $t$ have a distance from the detector smaller than the distance they cover during the interval $\de t$, that is $v\of t \, \de t$.}
\label{fig:BoltzmannCylinder}
\end{figure}

We can now repeat the derivation of section \ref{sec:EasyDerivation} using the Bohmian velocity instead of the quantum-mechanical momentum.
We consider again an initial state $\psi_0$ such that $\psi_0\of x \approx 0$ if $x\geq0$, but we do not ask anymore the momentum to be positive: we rather ask the Bohmian velocity to stay positive for every time after the initial state is prepared.
The particle crosses the detector between the times $t$ and $t+\de t$ if at time $t$ they are separated by a distance less than $v\of{x=0, t} \,\de t$ (cf.\ fig.\ \ref{fig:BoltzmannCylinder}).
The probability that at time $t$ the particle is in this region is $ v\of{0, t} \, |\psi\of{0, t}|^2\, \de t$, thus the probability density of arrival times is simply 
\begin{equation}
\Pi\of{t} =  v\of{0, t}\, |\psi\of{0, t}|^2 = j\of{0,t}.
\end{equation}

If  the velocity does not stay positive, it is still true that the particle crosses the detector during $(t, t+\de t)$ if at $t$ they are closer than $v\of{x=0, t} \,\de t$, but now this distance can also be negative.
In this case the current $j\of{0,t}$ still entails information about the crossing probability, but it also contains information about the direction of the crossing.
To get a probability distribution from the current we have to clearly specify how to handle the crossings from behind the detector and the multiple crossings of the same trajectory.
For example, one can count only the first time that every trajectory reaches the detector position, disregarding any further crossing, getting the so-called \emph{truncated current} \citep{DaumerDurrGoldstein1997,GrublRheinberger2002}.

The Bohmian analysis is readily generalized to three dimensions with an arbitrarily shaped detector, in which case also the arrival position is found.
More complicated situations, like the presence of a potential, or an explicit model for the detector, can be easily handled too.
Note that the presence of the detector can in principle be taken into account by use of the so-called \emph{conditional wave function} \citep{DuerrGoldsteinZanghi1992,PladevallOriolsMompart2012}, that allows to calculate the actual Bohmian arrival time in exactly the same way as described in this section, although the apparatus needs to be explicitly considered.

\subsection{Is the Bohmian arrival time measurable in an actual experiment?}

Any distribution calculated from the trajectories conveys some aspects of the actual motion of the Bohmian particle.
Such a distribution does not need in principle to have any connection with the results of a measurement, similarly to the Bohmian velocity that is not directly connected to the results of a momentum measurement.
The Bohmian level of the description is the one we should refer to when arguing about intrinsic properties of the system rather than measurement outcomes.
Since, in the framework of Bohmian mechanics, an intrinsic arrival time exists, namely that of the Bohmian particle, one should ask the intrinsic question that constitutes the title of this section rather than asking the apparatus dependent question
\question{When will the detector click?}
We do not mean that the latter question is irrelevant, to the contrary, it points towards the prediction of experimental results, that is of course of high value.
We shall continue the discussion of the latter topic in section \ref{sec:WhenClicks}.

\subsubsection{Linear measurement of the Bohmian arrival time}\label{sec:BohmPOVM}
We now ask if a linear measurement exists, such that its outcomes are the first arrival times of a Bohmian particle. 
For sure, this can not be exactly true, indeed, if this was the case, then the outcomes of such an experiment would be distributed according to the truncated current, that depends explicitly on the trajectories and is not sesquilinear with respect to the initial wave function as needed for a \povm{}.

However, it is reasonable to expect it to be approximately correct for some set of ``good'' wave functions.
That is motivated by the following considerations.
A typical position detector is characterized by a set of sensitive regions $\{ A_i \subset\R \}_{i=0,\dots,N}$, each triggering a different result.
If the measurement is performed at a fixed time $t$, and  if we get the answer $i$, then the Bohmian particle is at that time somewhere inside the region $A_i$.
A time measurement is usually performed with a very similar set up: one uses a position detector with just one sensitive region $A_0$ (in our case located around the origin) and waits until it fires.
In the ideal case, the reaction time of the detector is very small, and we can consider that the click occurs right after the Bohmian particle entered the sensitive region.
As a consequence, if the Bohmian trajectories cross the detector region only once and do not turn back in its vicinity, then we can expect the response of the actual detector to be very close to the quantum current.
This puts forward the set of wave functions such that the Bohmian velocity stays positive as a natural candidate for the set of good wave functions.
Surprisingly, it can be shown that \emph{there exists no \povm{} which approximates the Bohmian arrival time statistics on all functions in this set}  \citep{VonaHinrichsDurr2013}.

On the other hand, it is easy to see that the Bohmian arrival time is approximately given by a measurement of the momentum for all \emph{scattering states}, i.e.\ those states that reach the detector only after a very long time, so that they are well approximated by local plane waves.
Numerical evidence for a similar statement for the states with positive Bohmian velocity and high energy was also produced \citep{VonaHinrichsDurr2013}, but a precise determination of the set of good wave functions on which the Bohmian arrival time can be measured is still missing.

An explicit example of a model detector whose outcomes in appropriate conditions approximate the Bohmian arrival time can be found in \citep{DamboreneaEgusquizaHegerfeldt2002}.

\subsubsection{Nonlinear measurement}
An alternative to a linear measurement that directly detects the arrival time of a Bohmian particle is the reconstruction of its statistics from a set of measurements by a nonlinear procedure.

A first possibility in this direction starts by rewriting  the probability current \eqref{eq:J} as
\begin{equation}
 j\of{ x, t} =     
 \bra{ \psi_t}  \  
 	\tfrac{1}{2m}\,   \bigl(  \ket x \bra x \hat p + \hat p \ket x \bra x  \bigr)
\ \ket{\psi_t}  ,
\end{equation}
where $\hat p = -i\hbar\,\partial_x$ is the momentum operator.
The operator $\hat j  \coloneqq  \tfrac{1}{2m}\,   \bigl(  \ket x \bra x \hat p + \hat p \ket x \bra x  \bigr)$ is selfadjoint, therefore it could be possible to measure the current at the position $x$ and at time $t$ by measuring the average value at time $t$ of the operator $\hat j$.
Unfortunately, the operational meaning of this operator is unclear.

A viable solution is offered by weak measurements.
As showed by \citet{Wiseman2007}, it is possible to measure the Bohmian velocity, and therefore the current, by a sequence of two position measurements, the first weak and the second strong, used for postselection.
Wiseman's proposal has been implemented with small modifications in an experiment with photons%
\footnote{This experiment did not, of course, show the existence of a pointlike particle actually moving on the detected paths, but only the measurability of the Bohmian trajectories for a quantum system.}
\citep{KocsisBravermanRavets2011}.
A detailed analysis of the weak measurement of the Bohmian velocity and of the quantum current has been carried out by  \citet{TraversaAlbaredaDi-Ventra2012}.

It is worth noting that the weak measurement of the Bohmian velocity, if intended as a calibration of a non-linear measurement as explained in sec.\ \ref{sec:NotPOVMs}, gives rise to a genuine measurement, i.e.\ one whose outcome reveals the actual velocity possessed by the particle in that run
\citep{DurrGoldsteinZanghi2009}.

\section{When will the detector click?}\label{sec:WhenClicks}

We still have to answer the question we  posed at the beginning:
\question{When will the detector click?}

Surely, for any given experiment there is a \povm{} that describes the statistics of its outcomes.
Such an object will depend on the details of the specific physical system and of the measurement apparatus used for the experiment.
That is true not only for time measurements, but for any measurement, and for quantum mechanics as for Bohmian mechanics.
Yet, we can speak for example of the position measurement in general terms, with no reference to any specific setting, as it was disclosing  an intrinsic property of the system.
How can that be?

One can speak of the position measurement and of its \povm{} in general terms because a \povm{} happens to exist, that has all the symmetry properties expected for a position measurement and that does not depend on any external parameter.
That suggests that some kind of intrinsic position exists independently of the measurement details.
Recalling how the \povm{}s have been introduced in sec.\ \ref{sec:Povms}, it is readily clear that they inherently involve an external system (the apparatus) in addition to the system under consideration, and therefore they encode the results of an interaction rather than the values of an intrinsic property.
We also saw in sec.\ \ref{sec:EasyDerivation} how interpreting quantum-mechanical statistics as intrinsic objects leads to a mistake.
It is therefore very important to keep in mind that all \povm{}s describe the interaction with an apparatus.
Having this clear, it still makes sense to look for a \povm{} that does not explicitly depend on any external parameter, meaning with this simply that one does not want to give too much importance to the details of the apparatus.
Such a \povm{} may be regarded for example as the limiting element of a sequence of finer and finer devices, and it does not necessarily correspond to any realizable experiment.
Nevertheless, the fortunate circumstance that occurs for position measurements, for which such an idealized \povm{} exists, does not need to come about for all physical quantities one can think of.

For the arrival time it is possible to show that some \povm{}s exist that have the transformation properties expected for a time measurement \citep{Ludwig1983a}, but in three dimensions it is not possible to arrive at a unique expression in the general case, i.e.\ to something independent of any external parameter.
To do so, one needs to restrict the analysis to detectors shaped as infinite planes, or similarly to restrict the problem to one dimension 
\comment{Both Kijowski and Werner use the transverse momentum, so they need this hypothesis.}
\citetext{\citealp{Kijowski1974,Werner1986}; see also \citealp{Giannitrapani1997,EgusquizaMuga1999,MugaLeavens2000}}.%
In this case, for arrivals at the origin, one finds the \povm{} 
\begin{align}
&K \of{t_1, t_2} 
=
\sum_{\alpha=\pm 1}  \intdefde{t_1}{t_2}{T} \ket{T,\alpha} \bra{T,\alpha}  ,
\\
&\text{with}\quad
\braket{p | T,\alpha}
=
\sqrt{ \frac{|p|}{m h} }\ 
	\theta\of{\alpha p}\ 
	e^{ \frac{i p^2 T} {2m\hbar} }    ,
\end{align}
that corresponds to the probability density of an arrival at time $t$
\begin{equation}\label{eq:KijowskiProbDensity}
\sum_{\alpha=\pm 1}  
| \braket{t,\alpha | \psi_0} |^2 
=
\sum_{\alpha=\pm 1}  
\frac{1} {m h}\,
\left|  
\intdefde{0}{\alpha\infty}{p}
\smash{\sqrt{ |p| }}\ 
\braket{ p | \psi_t}
\right |^2   .
\end{equation}
Note that $K$ is not a projector valued measure because $\braket{T,+ | T,-}\neq0$.
For scattering states $K$ becomes proportional to the momentum operator, and the density \eqref{eq:KijowskiProbDensity} gets well approximated by the probability current \citep{Delgado1998}.
The general conditions under which this approximation holds are still not clear.

\subsection{The easy derivation, once again}

The analysis of sec.\ \ref{sec:BohmPOVM} of the measurability of the Bohmian arrival time translates quite easily in an approximate derivation of the response of a detector: essentially what we tried to do in sec.\ \ref{sec:EasyDerivation}, just right.

Consider again the setting described in sec.\ \ref{sec:EasyDerivation}, but with an initial state such that the Bohmian velocity stays positive.
That is equivalent to ask that the probability current stays positive, and therefore that the probability that the particle is on the left of the detector decreases monotonically in time.
As described in sec.\ \ref{sec:BohmPOVM}, thinking of the arrival time detector as of a position detector with only one sensitive region $A_0$ around the origin, it is reasonable to expect that for some set of good wave functions the detector will click right when the particle enters $A_0$.
Hence, the probability of a click at time $t$ is approximately equal to the increase of the probability that the particle is inside $A_0$ at that time, i.e.\ to the probability current through the detector.
Therefore, for the good wave functions, the probability current  is expected to be a good approximation of the statistics of the clicks of an arrival time detector.
As remarked in sec.\ \ref{sec:BohmPOVM} the set of the good wave functions is not exactly known, although it is clear that the scattering states are among its elements, and possibly also the states with positive probability current and high energy.

\section*{Acknowledgments}
The determination of the Bohmian trajectories shown in fig.\ \ref{fig:NegativeCurrent} is based on the code by Klaus von Bloh for the double slit available at \emph{\url{http://demonstrations.wolfram.com/CausalInterpretationOfTheDoubleSlitExperimentInQuantumTheory/}}.

Nicola Vona gratefully acknowledges the financial support of the Elite Network of Bavaria.

\newpage

\section*{Appendix: Example of Backflow}
\addcontentsline{toc}{section}{Appendix: Example of Backflow}

\begin{figure}
\begin{minipage}{\textwidth/2}%
\centering
\subfloat[\label{subfig:RhoTofX}]{\includegraphics[width=\textwidth]
{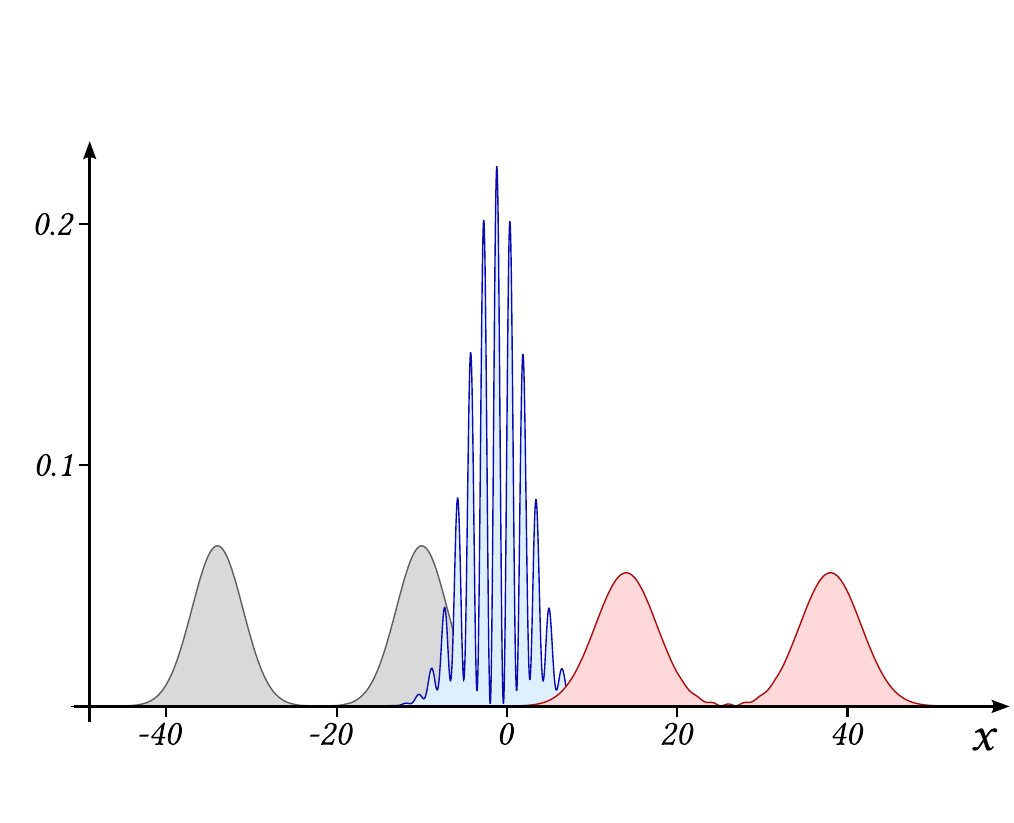}}%
\\
\subfloat[\label{subfig:RhoofTandX}]{\includegraphics[width=\textwidth]{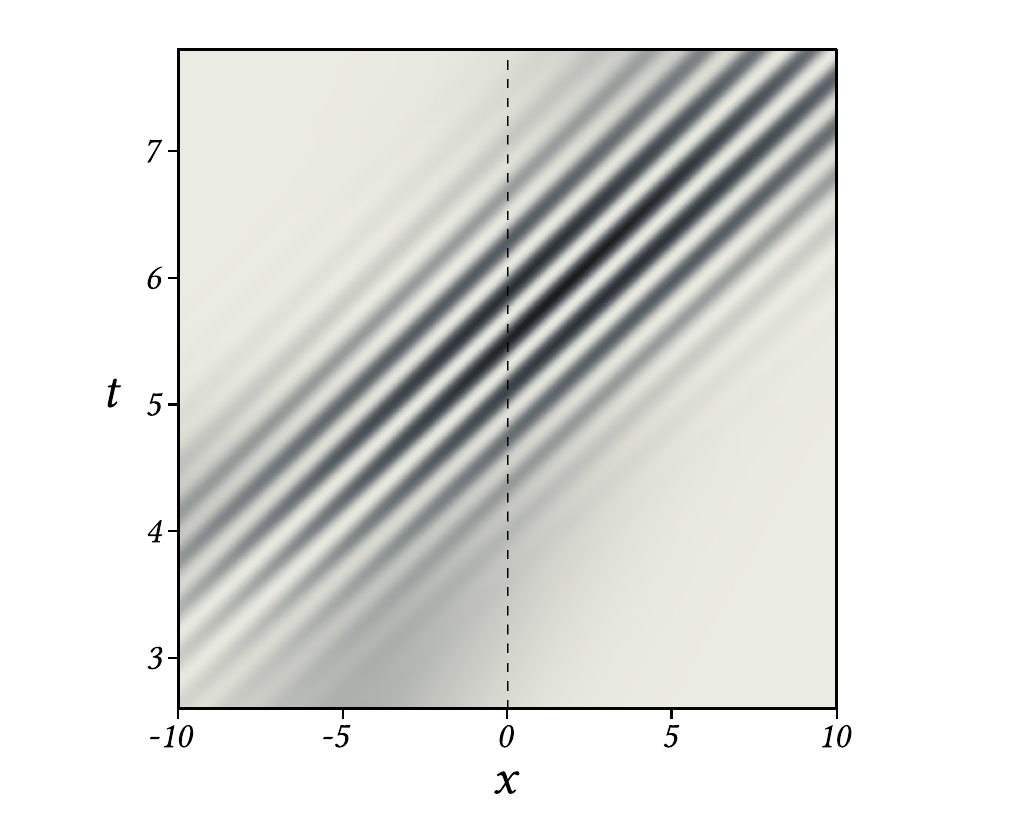}} 
\\
\subfloat[\label{subfig:CurrentAtScreen}]{\includegraphics[width=\textwidth]{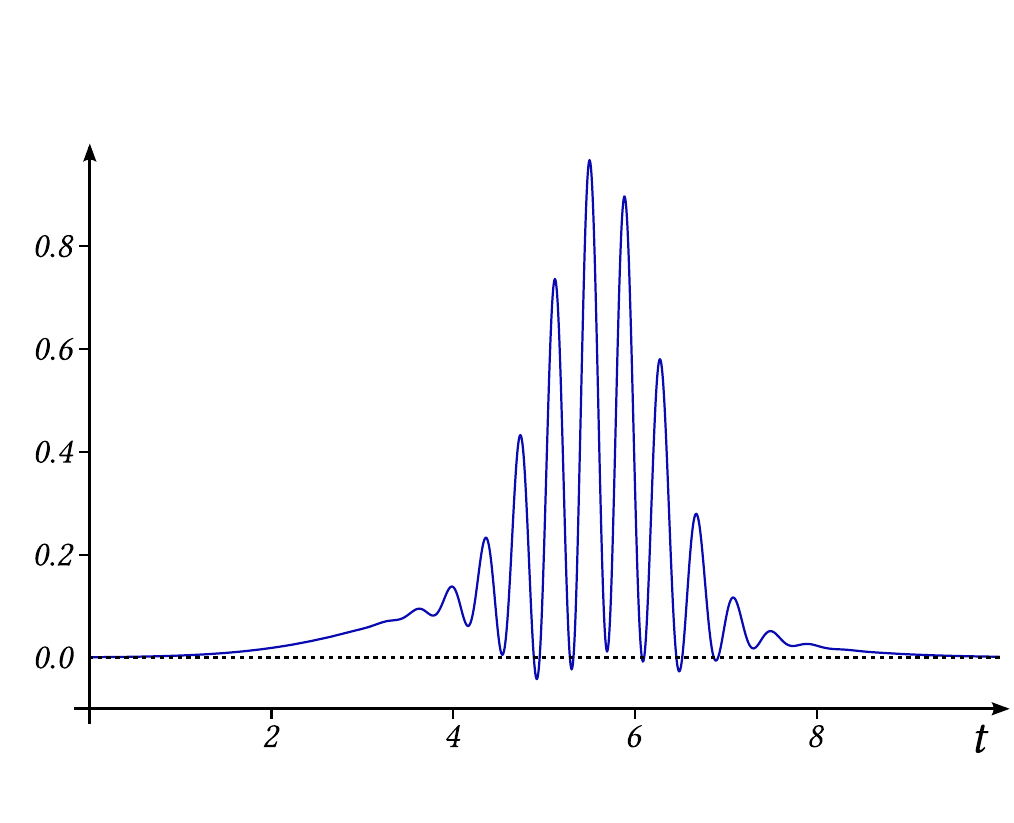}} 
\end{minipage}%
\centering
\begin{minipage}{\textwidth/2}%
\subfloat[\label{subfig:Traj}]{\includegraphics[width=\textwidth]{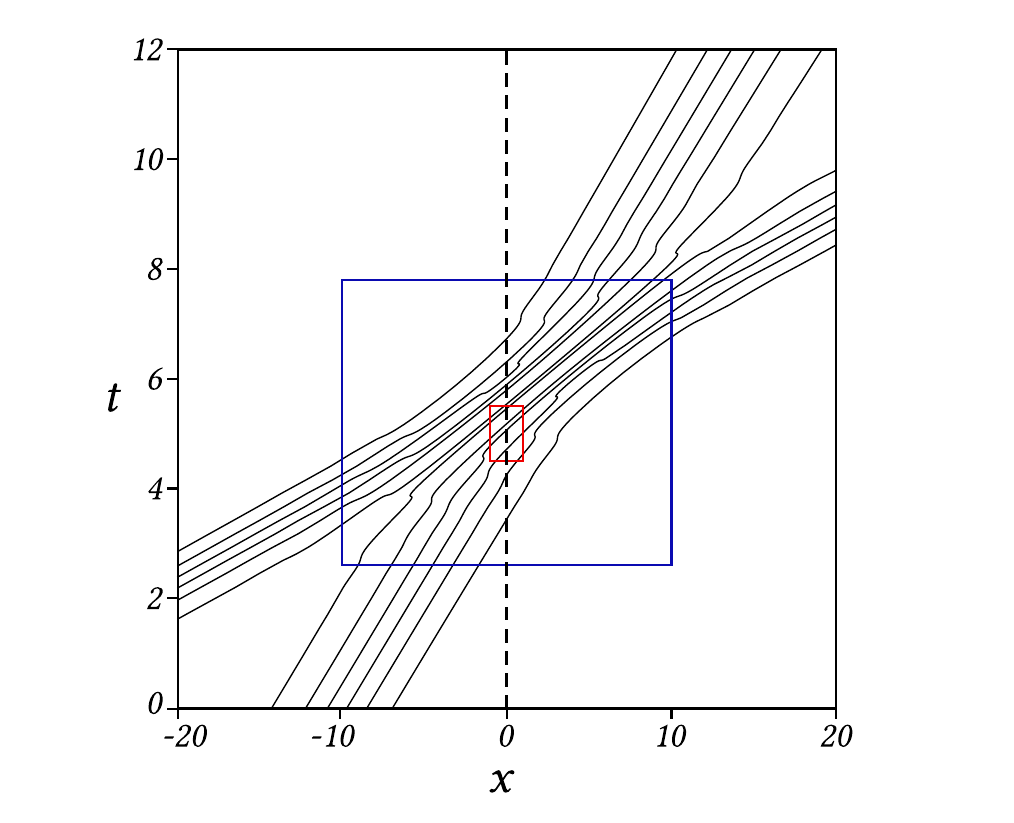}}%
\\
\subfloat[\label{subfig:TrajCloseUp}]{\includegraphics[width=\textwidth]{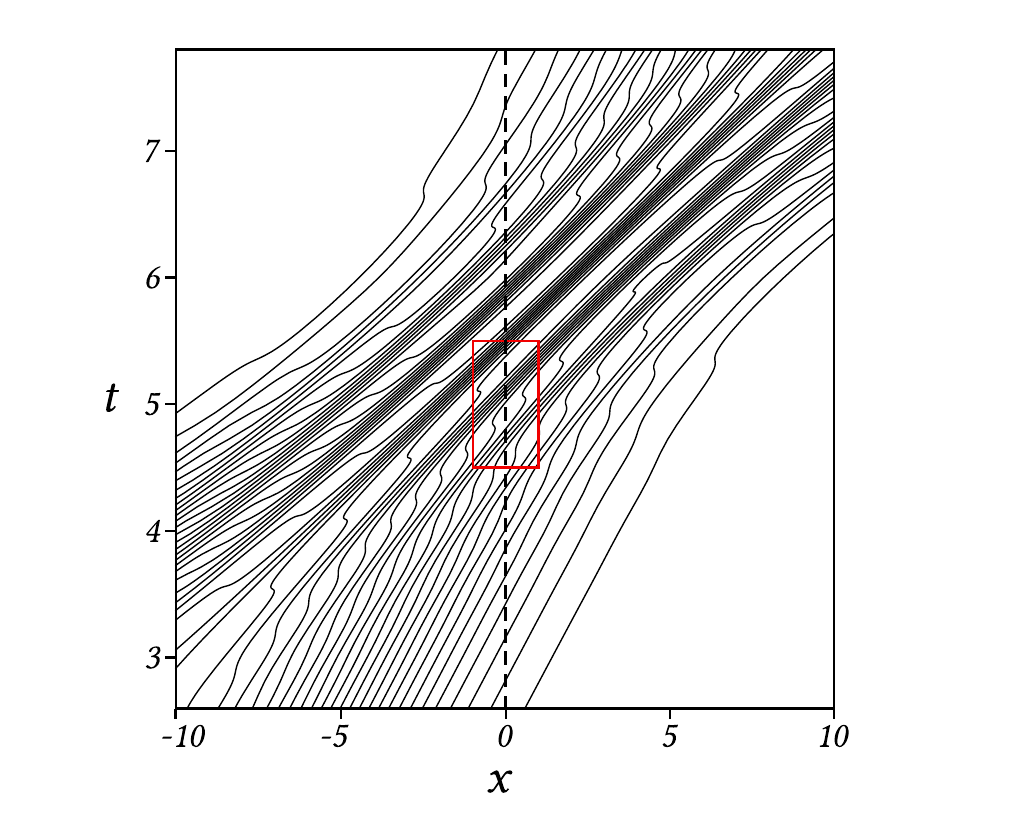}} 
\\
\subfloat[\label{subfig:TrajVeryCloseUp}]{\includegraphics[width=\textwidth]{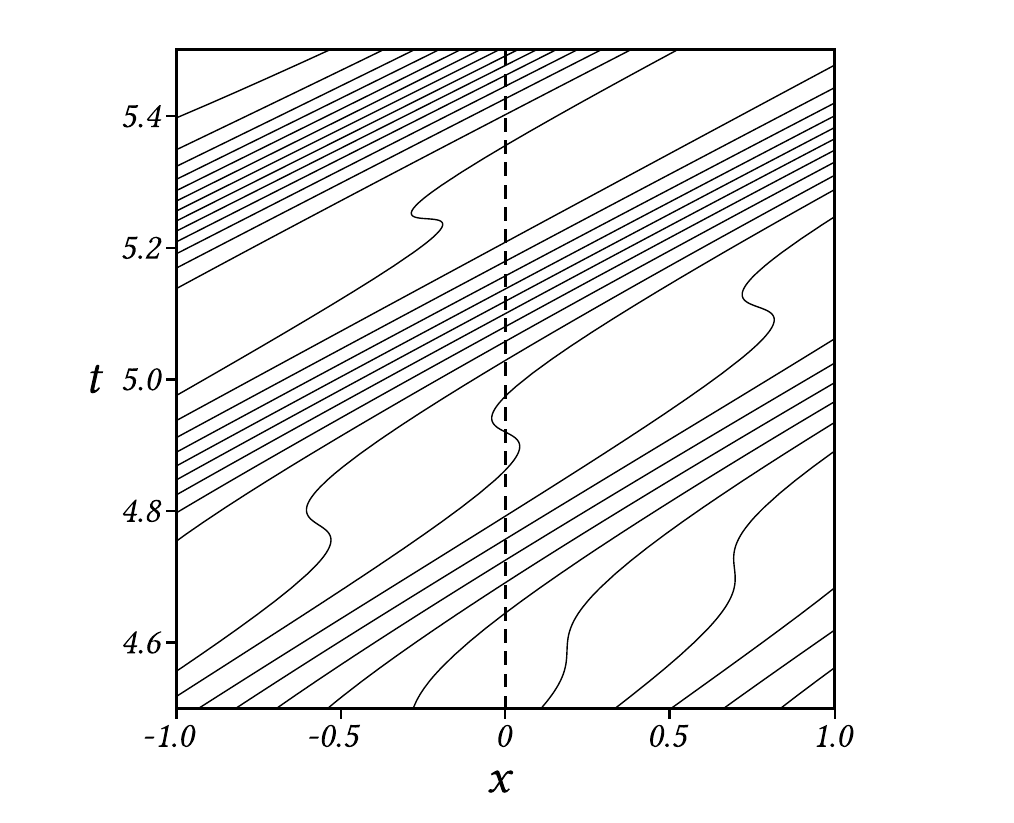}} 
\end{minipage}%
\caption[]{
Example of backflow: superposition of two gaussian packets (for the parameters see text).
The dashed line always represents the detector.
\subref{subfig:RhoTofX} 
Probability density of the position at time $t=0$ (gray), $t=5.2$ (blue), and $t=12$ (red).
\subref{subfig:RhoofTandX} 
Probability density of the position as a function of position and time. 
\subref{subfig:CurrentAtScreen} 
Probability current at the screen as a function of time. 
\subref{subfig:Traj} 
Overall structure of the Bohmian trajectories.
The blue and the red rectangles are magnified in \subref{subfig:TrajCloseUp} and \subref{subfig:TrajVeryCloseUp}.
}
\label{fig:NegativeCurrent}
\end{figure}

We mentioned that, even for states freely evolving and with support only on positive momenta, the quantum current can become negative.
We provide now a simple example of this circumstance, depicted in fig.\ \ref{fig:NegativeCurrent}.
We use units such that $\hbar=1$, and choose the mass to be one.

We consider the superposition of two gaussian packets, both with initial variance of position equal to $3$, corresponding to a variance of momentum of $1/6$.
The first packet is  initially centered in $x=-10$ and moves with average momentum $p=2$, while the second packet is centered in $x=-34$ and has  momentum $p=6$.
The probability of negative momentum is in this case negligible.
The second packet overcomes the first when they are both in the region around the origin, where the detector is placed.
In this area the two packets interfere, but then they separate again (cf.\ fig.\ \ref{subfig:RhoTofX}).

In fig.\ \ref{subfig:Traj} the Bohmian trajectories are shown on a big scale.
One can see that they never cross, but rather switch from one packet to the other. 
Moreover, they are almost straight lines, except for the interference region.
In that region, it is interesting to look at a higher number of trajectories, making apparent that the trajectories bunch together, resembling the interference fringes (cf.\ fig.\ \ref{subfig:RhoofTandX} and \ref{subfig:TrajCloseUp}).

Looking at the trajectories more in detail (fig.\ \ref{subfig:TrajVeryCloseUp}), one can see that they suddenly jump from one fringe to the next, somewhen even inverting the direction of their motion.
In this case, it can happen that the particle crosses the detector backwards, leading to a negative current, as shown in fig.\ \ref{subfig:CurrentAtScreen}.

One could argue that gaussian packets always entail negative momenta, and that this could be the cause of the negative current.
To show that this is not the case, we can compare the probability to have negative momentum 
\begin{equation}
\PP (p<0) = \intdef{-\infty}{0} |\tilde\psi\of p|^2 \, \de p \approx 10^{-33}
\end{equation}
with  the probability to have a negative Bohmian velocity 
\begin{equation}
\PP(v\of {t} <0) = \int_{K_t} \rho\of{x,t} \,\de x  ,
\end{equation}
where $K_t \coloneqq \{  x \in \R  |   j\of{ x, t}  < 0 \}$.
For instance, at time $t=5.2$ this probability is $0.008$ (numerically calculated), therefore the negative current can not be caused by the negative momenta.

\vfill
\phantomsection\addcontentsline{toc}{section}{References} 
\bibliography{biblio}

\bibliographystyle{plainnat.bst}
\end{document}